\documentclass[journal]{IEEEtran}
\pdfoutput=1 
\usepackage{graphicx} 
\usepackage{amsmath} 
\usepackage{amssymb}
\usepackage{cite}
\usepackage{amsfonts}
\usepackage{subfigure}
\usepackage{caption}
\usepackage{psfrag}
\usepackage{array}
\usepackage{multirow}
\usepackage{multicol}
\usepackage{mathtools, cuted}
\usepackage{setspace}
\usepackage{longtable}
\usepackage{float}
\usepackage{color}
\usepackage{algorithm}
\usepackage{algorithmic}
\usepackage[section]{placeins}

\newcommand{\bm}{\mathbf}

\newcommand{\be}{\begin{equation}}
\newcommand{\ee}{\end{equation}}
\newcommand{\bea}{\begin{eqnarray}}
\newcommand{\eea}{\end{eqnarray}}

\newcommand{\br}{{\bm r}}

\newcommand{\bA}{{\bm A}}
\newcommand{\bI}{{\bm I}}
\newcommand{\bW}{{\bm W}}

\newcommand{\bD}{{\bf D}}

\newcommand{\bH}{{\bf H}}
\newcommand{\bP}{{\bf P}}

\newcommand{\bg}{{\bf g}}

\newcommand{\bU}{{\bf U}}

\newcommand{\bd}{{\bf d}}

\newcommand{\bs}{{\bf s}}
\newcommand{\bx}{{\bf x}}

\newcommand{\bzero}{{\bf 0}}

\newcommand{\n}{{\bm n}}

\newcommand{\snri}{\mbox{$\frac{\sigma_n^2}{\sigma_d^2}$}}
\newcommand{\complex}{\mathbb{C}}
\renewcommand{\natural}{\mathbb{N}}

\newcommand{\delaymatrix}{\bm{\Pi}}
\newcommand{\dopplermatrix}{\bm{\Delta}}
\newcommand{\delayindex}{l}
\newcommand{\dopplerindex}{k}

\newcommand{\subcarrier}{\Delta f}
\newcommand{\kron}{\otimes}
\newcommand{\bn}{{\bf n}}
\DeclarePairedDelimiter{\ceil}{\lceil}{\rceil}
\DeclarePairedDelimiter{\floor}{\lfloor}{\rfloor}
\date{\today}

\begin{document}

\title{Circularly Pulse Shaped Orthogonal Time Frequency Space Modulation}
\author{Shashank Tiwari and Suvra Sekhar Das\\ G. S. Sanyal School of Telecommunications\\Indian Institute of Technology, Kharagpur}
\maketitle

\begin{abstract}
Orthogonal time-frequency space (OTFS) modulation is a recently proposed waveform for efficient data transfer in  high-speed vehicular scenario. Use of rectangular pulse shape in OTFS results in high out of band (OoB) radiation, which is undesirable for multi-user scenarios. In this work, we present a circular pulse shaping framework for OTFS for reducing the OoB. We also design a low complexity transmitter for such a system. We argue in favor of orthogonal transmission for low complexity transceiver structure. We establish that frequency-localized circulant Dirichlet pulse is one of the possible pulses having this desirable unitary property, which can reduce OoB radiation significantly (by around 50 dB) without any loss in BER. We also show that our proposed pulse shaped OTFS has a lower peak to average power ratio than conventional OTFS system.
\end{abstract}

\section{Introduction}

Vehicular channels are in general time-varying due to significant values of delay and Doppler spreads \cite{matolak_air-ground_2017}. Hadani et al. proposed orthogonal time-frequency space (OTFS) modulation \cite{hadani_orthogonal_2017} to provide communication in such scenarios by modulating data symbols in the delay-Doppler domain, which makes it approximately invariant to a time-varying channel. OTFS modulation has two stages; in the first stage, delay-Doppler domain data symbols are mapped to dual time-frequency data symbols using inverse symplectic fast Fourier transform (ISFFT). In the second stage, time-frequency symbols are converted to the time domain using time-frequency modulator, which is usually an orthogonal frequency division (OFDM) modulator. In this work, we propose a more generalized modulator in the second stage to enable changes in transmitter properties such as out of band (OoB) radiation, peak to average power ratio (PAPR), etc.

Conventional OTFS system as in \cite{hadani_orthogonal_2017,raviteja_interference_2018,farhang_low_2017} uses one symbol long rectangular pulse shape. It is known that rectangular pulse has very high OoB radiation. In this work, we demonstrate that OTFS with rectangular pulse shape has high OoB radiation, which can result in high adjacent channel interference. To reduce OoB radiation in multi-carrier communication, use of frequency localized pulse shapes are recommended \cite{banelli_modulation_2014} instead of a rectangular pulse. To the best of our knowledge, the first attempt to include pulse shaping in OTFS was made in \cite{raviteja_practical_2018}, but, pulse shapes considered therein \cite{raviteja_practical_2018} are one symbol long. They induce non-orthogonality which degrades BER performance. 
Authors in \cite{surabhi_peak--average_2019}  analyzed the effect of pulse shaping on the PAPR performance of OTFS system and found that Gaussian, as well as raised cosine pulse shaping, increases the PAPR of the OTFS system.  
 In this work, we are interested in pulse shapes which, despite making OTFS more frequency localized, does not introduce any non-orthogonality, hence do not degrade BER performance. We also analyze the effect of pulse shaping on PAPR performance.

A linear pulse reduces spectral efficiency as it spreads out of the symbol which motivates us to consider circular pulse shape in order to retain spectral efficiency of OTFS system. Thus, we propose a circular pulse shaped (CPS) framework for OTFS in which pulses are circular and can span the whole frame duration. CPS-OTFS transmission can be written as matrix-vector multiplication, which requires a quadratic order of complexity. We design a low complexity CPS-OTFS transmitter having log-linear order of complexity using the matrix factorization of matrices involved in transmission.

Further, we investigate circular pulse shapes which have the following two properties, (1) Unitary, i.e., pulse shapes for which OTFS system becomes unitary and (2) frequency-localized which can reduce the OoB radiation. We show that circulant Dirichlet pulse shaped OTFS (CDPS-OTFS) has these two above-mentioned properties. We further demonstrate that  CDPS-OTFS can achieve significant OoB radiation reduction without any loss in BER. We also show that PAPR of CDPS-OTFS is lower than conventional OTFS systems.

 We use the following notations throughout the paper. We let $\bx$, $\bm X$ and $x$ represent vectors, matrices and scalars respectively. The superscripts $(.)^{\rm T}$and $(.)^{\rm H}$ indicate transpose and conjugate transpose  operations, respectively. $\bI_N$ and $\bW_L$ represents identity matrix with order $N$ and $L$-order normalized IDFT matrix respectively. Kronecker product operator is given by $\kron$. 
$diag\{.\}$ is a diagonal matrix whose diagonal elements are formed by the elements of the vector inside.
$circ\{.\}$ is a circulant matrix whose first column is given by the vector inside. $E\{.\}$ is exception of expression. $\ceil{-}$ and $\floor{-}$ are  ceil and floor operators respectively. $\natural$ represents natural numbers.  $j=\sqrt{-1}$.

\section{Circular pulse shaped OTFS (CPS-OTFS)} 
\begin{figure}[h]
	\centering
	\includegraphics[width=1\linewidth]{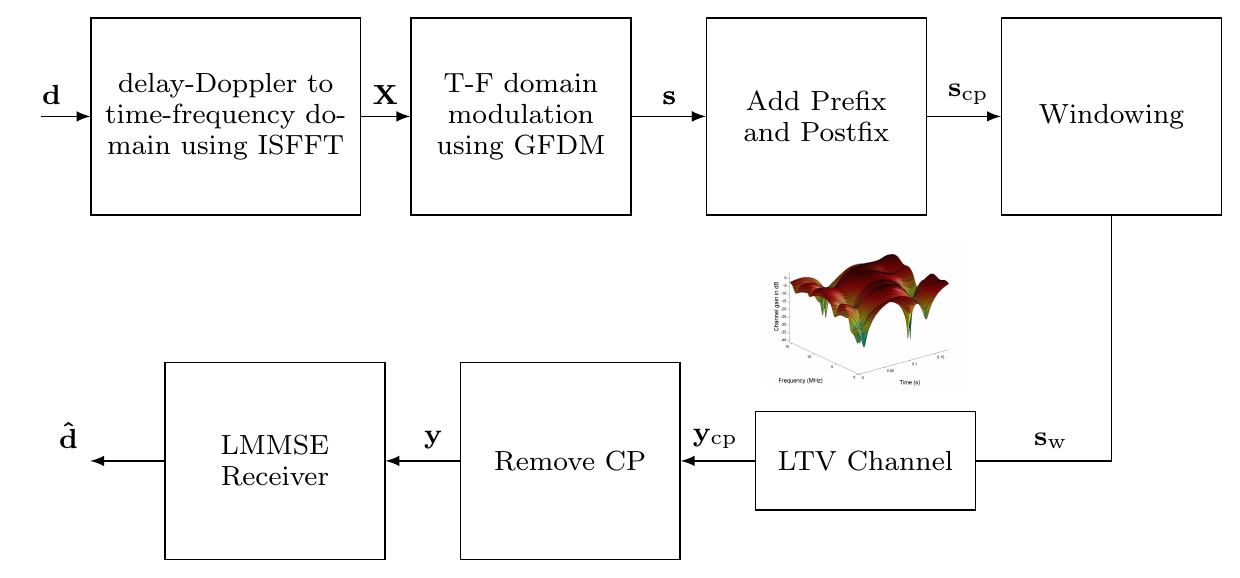}
	\caption{Circular Pulse Shaped OTFS System}
	\label{fig:ddsgfdmblockdiagram}
\end{figure}
CPS-OTFS system can be understood in the light of Fig.~\ref{fig:ddsgfdmblockdiagram}. We consider an OTFS system with $T_f=NT$ frame duration and $B=M\subcarrier$ bandwidth having $N$ number of time-symbols with $T$ symbol duration and $M$ number of sub-carriers with  $\subcarrier$ bandwidth.
QAM modulated data symbols, $d(k,l)\in \complex$, $k\in \natural[0~N-1]$, $l\in \natural[0~M-1]$, are arranged over Doppler-delay lattice $\Lambda=\{(\frac{k}{NT},~\frac{l}{M \subcarrier})\}$. We assume that data symbols are independent and identical i.e. $E[d_{k,l} d_{k',l'}^\ast] = \sigma_d^2 \delta_{k-k',l-l'} $. Doppler-delay domain data $d(k,l)$ is mapped to time-frequency domain data $X(n,m)$  on lattice $\Lambda^{\perp}=\{(nT,~m\subcarrier)\}$, $n\in \natural[0~N-1]$ and $m\in \natural [0~M-1]$ by using inverse symplectic fast Fourier transform (ISFFT). $X(n,m)$ can be given as \cite{hadani_orthogonal_2017},
\begin{equation}
X(n,m)=\frac{1}{\sqrt{NM}} \sum_{k=0}^{N-1}{\sum_{l=0}^{M-1}{d(k,l) e^{j2\pi [\frac{nk}{N}-\frac{ml}{M}]}}}.
\label{eqn:ISFT}
\end{equation}
Each time-frequency domain data $X(n,m)$ is pulse shaped using circular pulse shape  $g_{(n,m)}(t)=g(t-mT)_{T_f} e^{j2\pi n\subcarrier (t-mT)}$, where $g(t)$ is a prototype pulse shape having length $T_f$. Time domain transmitted signal is obtained as,
\begin{equation}
s(t)= \sum_{n=0}^{N-1}{\sum_{m=0}^{M-1}{X(n,m) } g((t-mT) \mod T_f) e^{j2\pi n\subcarrier (t-mT)}}
\label{eqn:GFDMMod}
\end{equation}
CPS-OTFS can also be viewed as an ISFFT precoded generalized frequency division multiplexing (GFDM) \cite{michailow_generalized_2014} system where time-frequency modulation is performed as GFDM modulation. CPS-OTFS system converges to pulse shaped OTFS in \cite{raviteja_practical_2018} when, $g(t)= \begin{cases}
a(t) ~ \text{if} ~ 0<t<T \\
0 ~ \text{otherwise}
\end{cases}$,
where $a(t)$ is a $T$ duration pulse shape. Rectangular pulse shaped OTFS (RPS-OTFS) as in \cite{hadani_orthogonal_2017,raviteja_interference_2018,farhang_low_2017} can be obtained by setting $a(t)=1$. After taking samples at $\frac{T}{M}$ sampling duration, transmitted signal is given as,
\begin{equation}
s(r)=\sum_{n=0}^{N-1}{\sum_{m=0}^{M-1}{X(n,m)g((r-nM)\mod MN) e^{j2\pi \frac{m r}{M}}}},
\end{equation}
for $r=\natural[0~MN-1]$.

We collect values of $d(k,l)$ in a vector as $\bd=[d(0,0)~d(1,0)~\cdots d(N-1,0)~ d(0,1)~d(1,1)~\cdots d(N-1,1)~\cdots d(0,M-1)~d(1,M-1)~\cdots d(N-1,M-1) ]^{\rm T}$ and similarly we take values of $X(n,m)$ is a vector as,  $\bx=[X(0,0)~X(1,0)~\cdots X(N-1,0)~ X(0,1)~X(1,1)~\cdots X(N-1,1)~\cdots X(0,M-1)~X(1,M-1)~\cdots X(N-1,M-1) ]^{\rm T}$. Using (\ref{eqn:ISFT}), $\bx$ is given as,
\begin{equation}
\bx=\underbrace{\bU_M^{\rm H} \bP \bU_N}_{\bA_{\rm DD}} \bd,
\end{equation}
where, $\bU_N=\bI_M \kron \bW_N$, $\bU_M=\bI_N \kron \bW_M$ and $\bP$ is a permutation matrix whose elements can be given as, $\bP(s,q)=\begin{cases}
1 ~\text{if}~ q= (s\mod M)(N-1)+\floor{\frac{s}{M}}\\
0 ~ \text{Otherwise}.
\end{cases}$, for $q,s \in \natural[0~MN-1]$. Further, collecting samples $s(r)$ as  $\bs=[s(0)~s(1) \cdots s(MN-1)]$, $\bs$ can be given as, $\bs=\bA_g \bx$, where $\bA_g$ is  GFDM modulation matrix. Thus, transmitted signal can be given in matrix-vector form as, 
\begin{equation}
\bs=\bA \bd, 
\label{eqn:otfsmodmatrix}
\end{equation}
where, $\bA=\bA_g \bA_{\rm DD}$ is CPS-OTFS modulation matrix. 
A cyclic prefix (CP) of length $\alpha' \geq \alpha-1$ is appended at the end and beginning of $s(r)$, $\alpha$ is channel delay length i.e. $\bs_{cp}=[\bs(MN-\alpha'+1:MN-1) ~ \bs^{\rm T} ~ \bs(0:\alpha'-1)]$.  Finally, $\bs_{cp}$ is multiplied to a widow of length $MN+2\alpha'$ which has values of one for middle $MN$ samples and soft edges for initial and last $\alpha'$ samples. 

We consider a time varying channel with $P$ paths having $h_p$ complex attenuation, $\tau_p$ delay and $\nu_p$ Doppler value for $p^{\rm th}$ path where $p\in \natural[1~P]$. Delay-Doppler channel spreading function can be given as,
\begin{equation}
h(\tau,\nu)=\sum_{p=1}^{P}{h_p \delta(\tau-\tau_p) \delta(\nu-\nu_p)}.
\end{equation}
The delay and Doppler values for $p^{\rm th}$ path is given as $\tau_p=\frac{\delayindex_p}{M\subcarrier}$ and $\nu_p=\frac{\dopplerindex_p}{NT}$.  We assume that $N$ and $M$ are sufficiently large so that there is no effect of fractional delay and Doppler on the performance. We also assume the perfect knowledge of $(h_p,~\delayindex_p,~\dopplerindex_p)$, $p\in \natural[0~P-1]$, at the receiver. Let $\tau_{max}$ and $\nu_{max}$ be the maximum delay and Doppler spread. Channel delay length $\alpha= \ceil{\tau_{max}M\subcarrier}$ and channel Doppler length, $\beta= \ceil{\nu_{max} NT}$.

After removal of CP  at the receiver, received signal can be written as \cite{raviteja_practical_2018},
\begin{equation}
\br=\bH \bs +\bn,
\end{equation}
where, $\bn$ is white Gaussian noise vector of length $MN$ with elemental variance $\sigma_\n^2$ and $\bH$ is a  $MN\times MN$ channel matrix given as,
\begin{equation}
\bH=\sum_{p=1}^{P}{h_p \delaymatrix^{\delayindex_p} \dopplermatrix^{\dopplerindex_p}},
\end{equation}
with $\delaymatrix=circ\{[0~1~0 \cdots 0]_{MN\times 1}\}$ is a circulant delay matrix and $\dopplermatrix=diag\{1~ e^{j2\pi\frac{1}{MN}}~ \cdots e^{j2\pi\frac{MN-1}{MN}}\}$ is a diagonal Doppler matrix. In this work, we consider linear minimum mean square error (LMMSE) receiver \cite{surabhi_diversity_2019} for its known interference cancellation capabilities \cite{kay1993fundamentals}.  
After processing $\br$ through a LMMSE equalizer, we get estimated data vector,
\begin{equation}
\hat{\bd}=(\bH\bA)^{\rm H} [(\bH \bA)(\bH \bA)^{\rm H}+\snri \bI]^{-1} \br. 
\label{eqn:MMSEOTFS}
\end{equation}

\section{Low Complexity Transmitter} 
CPS-OTFS modulation using (\ref{eqn:otfsmodmatrix}) requires complex multiplications (CMs) in the $O(M^2 N^2)$ which can be computationally a burden when values of $MN$ are high. Here we present a low complexity transmitter for practical implementation of CPS-OTFS. Using the factorization of $\bA_g$ given in \cite{tiwari_low-complexity_2018}, (\ref{eqn:otfsmodmatrix}) can be given as,
\begin{eqnarray}
\bs&=& \underbrace{\bP \bU_{N} \bD \bU_N^{\rm H} \bP^{\rm T} \bU_{M}}_{\bA_g} \underbrace{\bU_M^{\rm H} \bP \bU_{N}}_{\bA_{\rm DD}} \bd \\
\label{eqn:lowcomplexitytransmitter}
&=& \underbrace{\bP \bU_{N} \bD}_{\bA} \bd,
\end{eqnarray}
where, ${\bD} = diag\{{\lambda}(0),~{\lambda}(1)\cdots {\lambda}(MN-1)\}$ is diagonal matrix, whose $r^{\rm th}$ element can be given as,
\begin{equation}
\lambda(r)=\sum_{m=0}^{N-1}{g[mM+\floor{\frac{r}{N}}] e^{j2 \pi \frac{m(r \mod N)}{M}}}
\end{equation}
Thus, using (\ref{eqn:lowcomplexitytransmitter}) and assuming that $\bD$ is computed off-line,  CPS-OTFS can be implemented using $M$ number of $N$-point IFFTs and $MN$-point scalar multiplier. Our proposed transmitter for CPS-OTFS requires $MN+\frac{MN}{2}\log_2{N}$ CMs which needs only $MN$ more CMs than RPS-OTFS in \cite{farhang_low_2017}.
 

\section{Circular Dirichlet pulse shaped OTFS (CDPS-OTFS)}
\begin{figure}[h]
	\centering
	\subfigure[Time Domain]{%
		\label{fig:first}%
		\includegraphics[width=0.46\linewidth]{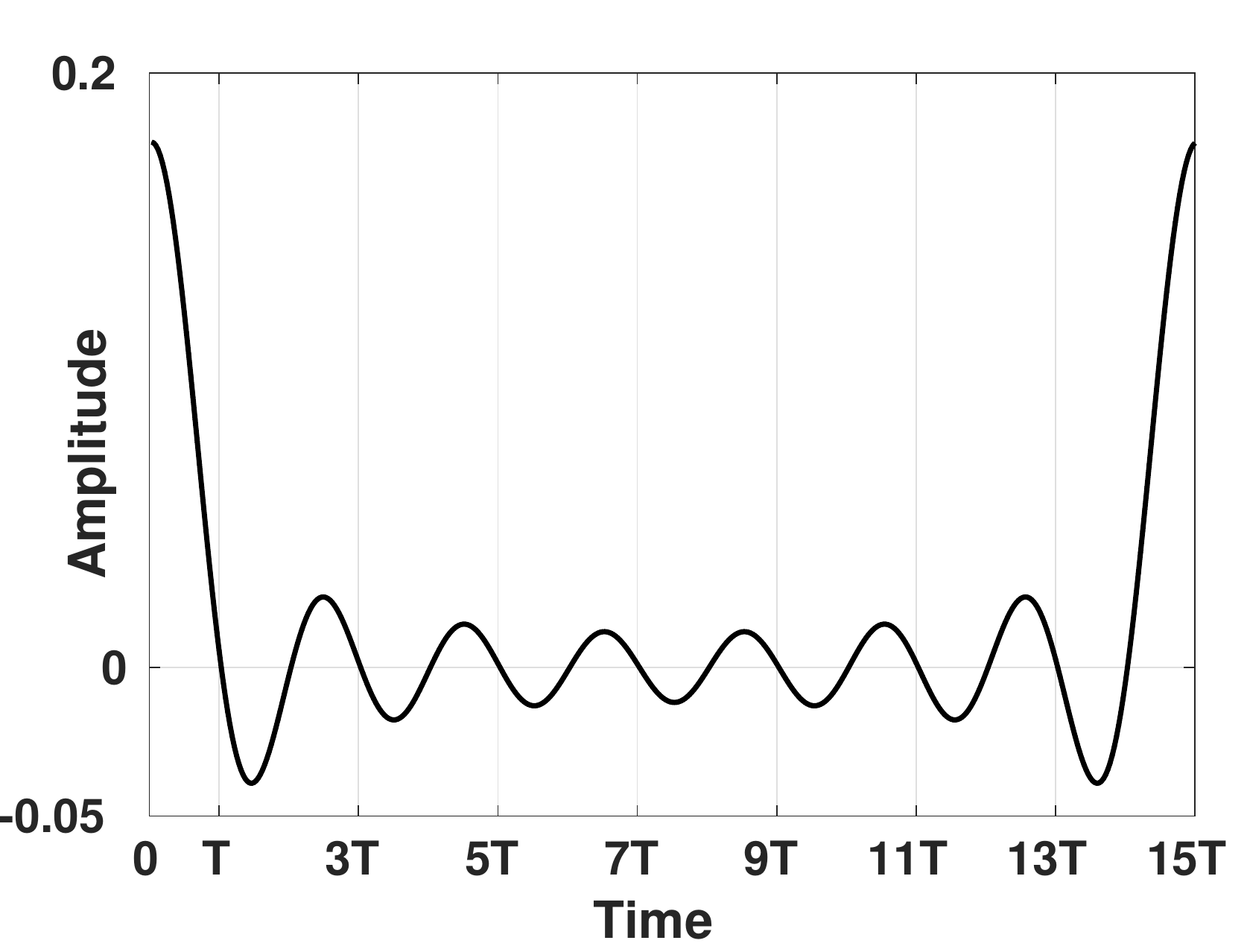}}%
	\qquad
	\subfigure[Frequency Domain]{%
		\label{fig:second}%
		\includegraphics[width=0.46\linewidth]{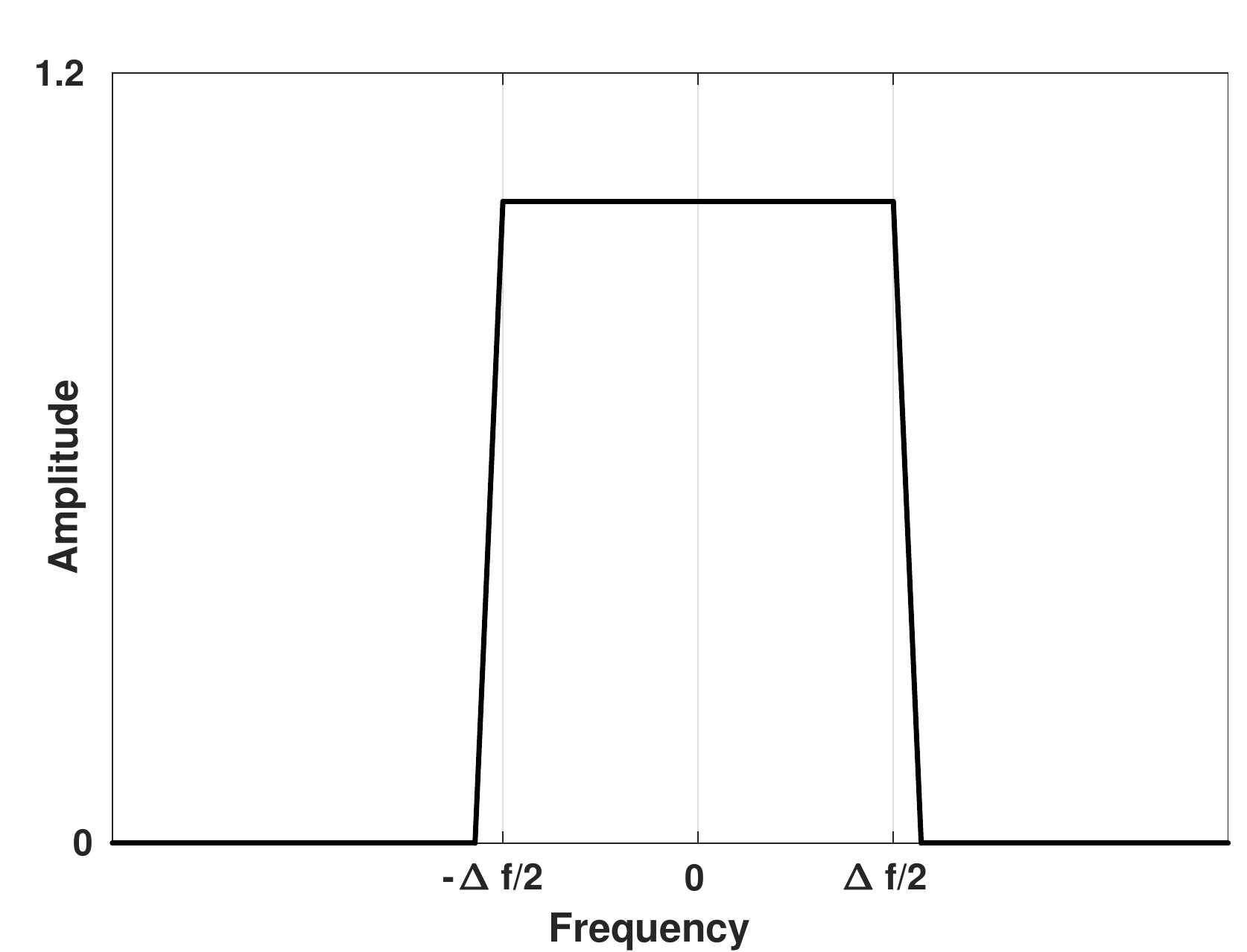}}%
	\caption{Illustration of frequency localization of Dirichlet Pulse for $N=15$ and $M=32$.}
	\label{fig:drichletpulse}
\end{figure}

 For unitary $\bA$ in (\ref{eqn:lowcomplexitytransmitter}), $\bD$ should be unitary as $\bP \bU_N$ are unitary i.e. $\bD \bD^{\rm H}=\bI_{MN}$ or $abs\{\bD\}=\bI_{MN}$. Trivial solution to this is a rectangular pulse for which $\bD=\bI_{MN}$. A class of prototype pulses for which $abs\{\bD\}=\bI_{MN}$ are called constant magnitude characteristics matrix (CMCM) pulses \cite{chen_matrix_2017}. Drichlet pulses are CMCM pulses which can be obtained by taking inverse Fourier transform of rectangular pulse. Let $\bg_{rect}=\frac{1}{\sqrt{M}}[1~1 \cdots 1~ \bzero_{MN-M}]^{\rm T}_{MN}$ be a rectangular pulse, Dirichlet pulse can be found by $\bg=\bW_{MN}^{\rm H} \bg_{rect}$ whose time and frequency domain shape can be understood by Fig.~\ref{fig:first} and \ref{fig:second} respectively. As opposed to rectangular pulses, Drichlet pulses are localized in frequency which helps in reducing OoB radiation. 
\section{Remarks on Receiver Complexity}
Since $\bA$ for RPS-OTFS as well as CDPS-OTFS is unitary, (\ref{eqn:MMSEOTFS}) is simplified to
\begin{equation}
\hat{\bd}=\underbrace{\bA^{\rm H}}_{MF} \underbrace{\bH^{\rm H}[\bH\bH^{\rm H}+\snri \bI]^{-1}}_{MMSE-CE} \br.
\label{eqn:MMSEOTFStwostage}
\end{equation}
Thus OTFS-MMSE receiver for CDPS-OTFS as well as RPS-OTFS is simplified to a two-stage receiver. The first stage is MMSE channel equalization (MMSE-CE), followed by a matched filter (MF) for OTFS. It is straightforward to see that MF operation for CDPS-OTFS requires additional $MN$-point scalar multiplications for multiplication of diagonal values of $\bD^{\rm H}$ as compared to RPS-OTFS. It can be concluded that implementation of MMSE receiver for CDPS-OTFS is only slightly higher than that of MMSE receiver for RPS-OTFS. 
\section{Simulation Results and Discussion}

\begin{table} [h]
	\centering
	\caption{Simulation Parameters}
	\label{tab:simu:para:mmse}
	\begin {tabular}{|m{0.45\linewidth}|m{0.45\linewidth}|}
	\hline
	Number of Sub-carriers $M$ & 512\\ \hline
	Number of Time-slots $N$ & 127 \\ \hline
	Window & Mayer root raised cosine with roll of factor =1 \cite{zayani_wola-ofdm:_2016}\\ \hline 
	Mapping & 4 QAM \\ \hline
	Sub-carrier Bandwidth & 15 KHz \\ \hline
	Channel & Extended Vehicular-A (EVA) \cite{series2009guidelines} \\ \hline
	Vehicular Speed (in Kmph)& 500   \\ \hline
	Carrier Frequency & 4 GHz \\ \hline
	Gaurd (null) sub-carriers for OoB computation & $\natural[1~128] \bigcup \natural[384~512]$ \\ \hline
	Circular prefix (CP) or post-fix value, $\alpha'$ & 64 \\ \hline
\end{tabular}
\end{table}

\begin{figure}[h]
	\centering
	\includegraphics[width=1\linewidth]{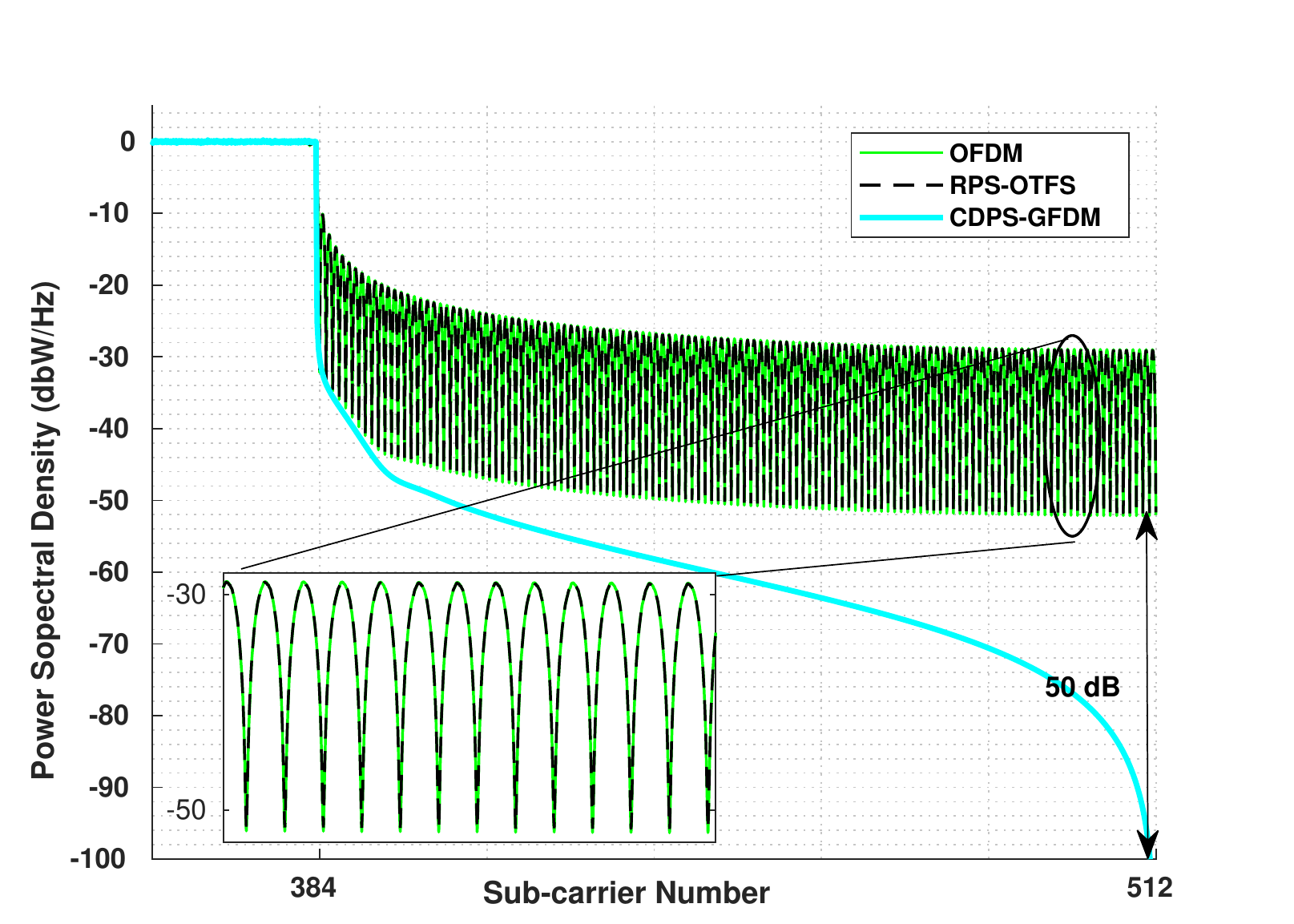}
	\caption{Comparison of Out of Band leakage for different waveforms.  }
	\label{fig:otfsspectrum}
\end{figure}
We demonstrate the OoB performance of our proposed CDPS-OTFS as well as that of RPS-OTFS through Monte-Carlo simulations.  We consider an OTFS with parameters as in Tab.~\ref{tab:simu:para:mmse}.  Power spectral density is plotted in Fig.~\ref{fig:otfsspectrum} for different waveforms. OFDM herein is a block OFDM having $M$ sub-carriers and $N$ time-slots. It can be observed that RPS-OTFS has similar PSD as OFDM. Side lobes of RPS-OTFS are found to fluctuate between -30 dB to -50 dB. Thanks to frequency localization provided by Dirichlet pulses, CDPS-OTFS has much lower side-lobes than RPS-OTFS. We can see that at the edge sub-carrier,  CPS-OTFS has nearly 50 dB lower OoB than RPS-OTFS.

\begin{figure}[h]
	\centering
	\includegraphics[width=1\linewidth]{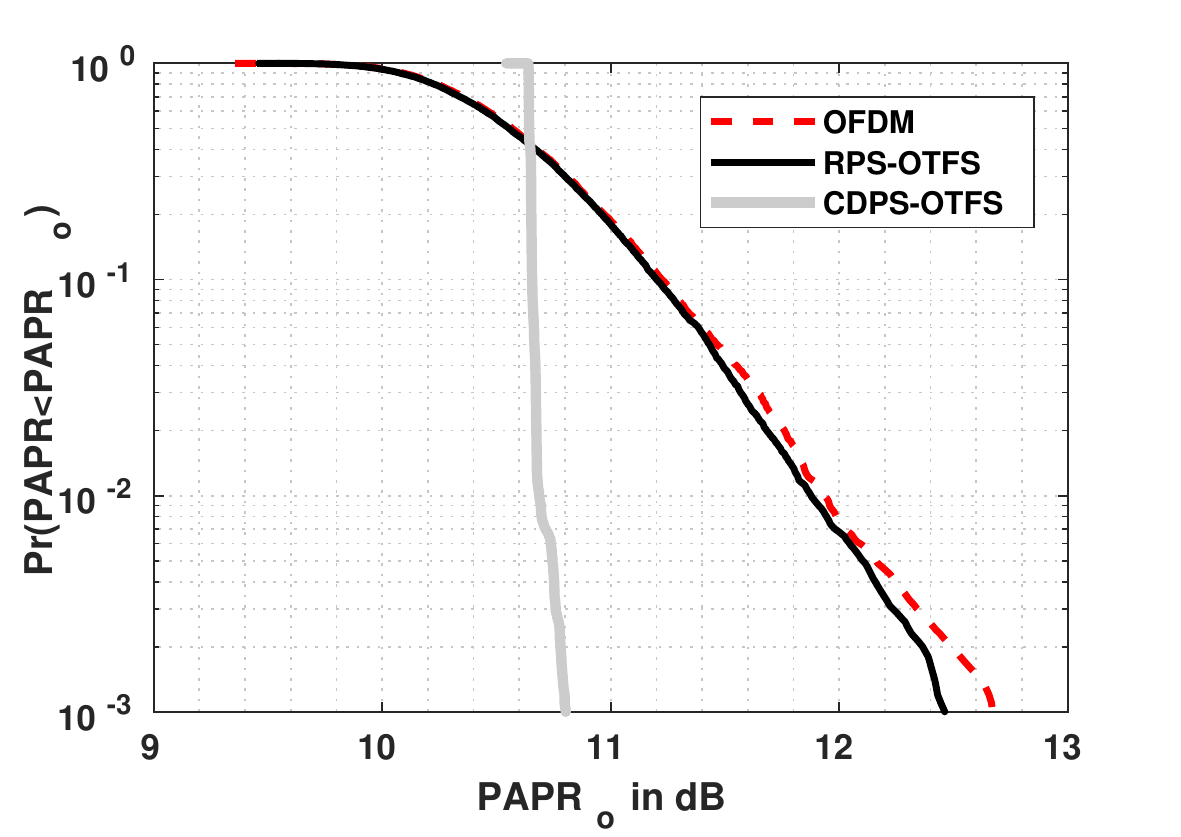}
	\caption{PAPR comparison of RPS-OTFS and CDPS-OTFS.}
	\label{fig:paprotfs}
\end{figure}

Next, we see the effect of circulant Drichlet pulse shaping on the PAPR. We compute PAPR for $10^5$ independent transmissions and plot complementary cumulative density function (CCDF) of the different waveforms in Fig.~\ref{fig:paprotfs}. It can be observed that RPS-OTFS has PAPR whose values are close to that of the PAPR values of OFDM. This follows the trend, for large values of N (in this work, N = 128 ), as shown in \cite{surabhi_peak--average_2019}.  Interestingly, Dirichlet pulse shaping reduces the PAPR as compared to rectangular pulse shaping in OTFS. When we compare the PAPR value that is exceeded with probability less than $10^{-3}$,  CDPS-OTFS has about 1.6 dB gain over RPS-OTFS.

\begin{figure}[h!]
	\centering
	\includegraphics[width=\linewidth]{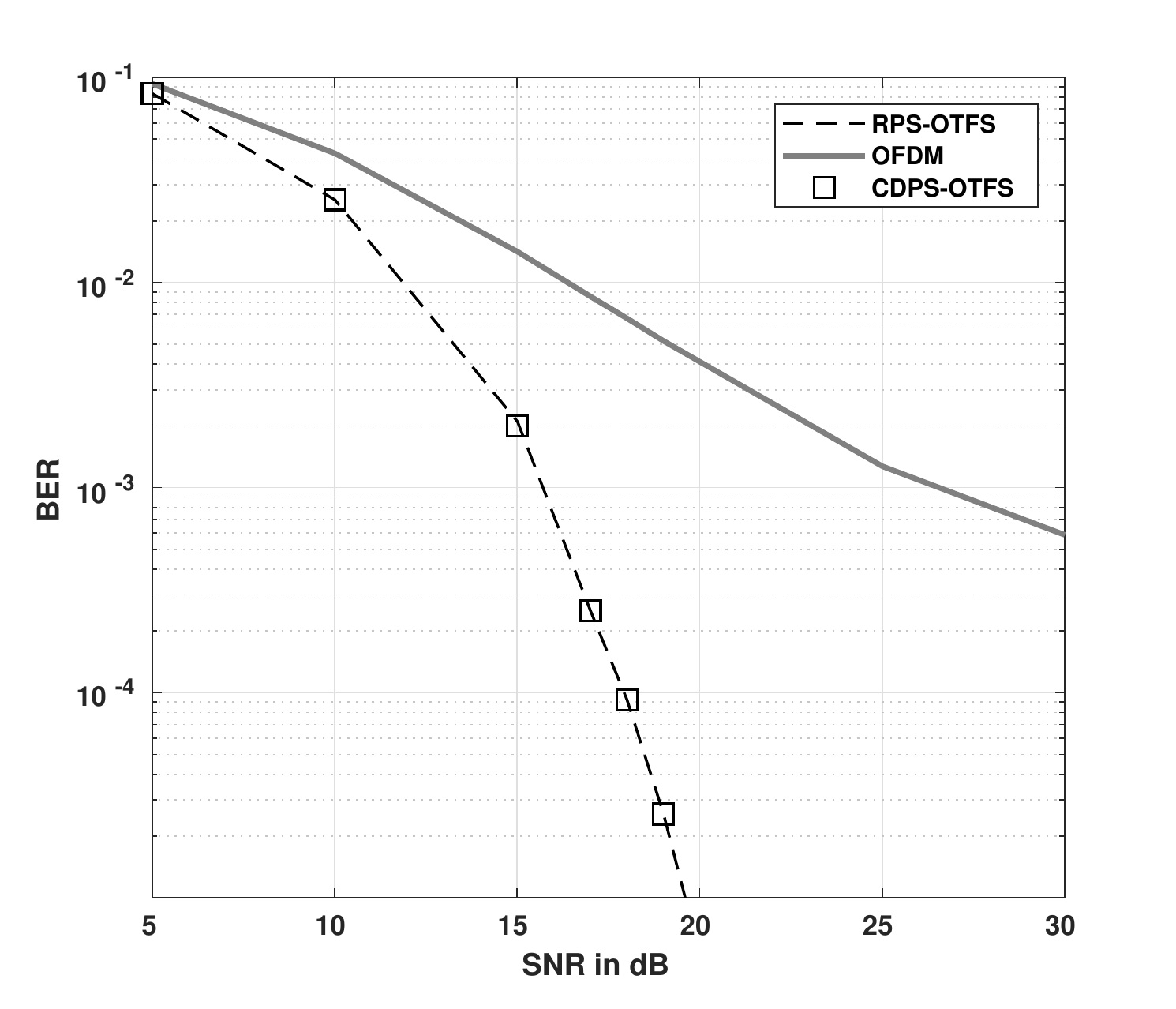}
	\caption{The BER performance of different waveforms in vehicular scenarios.}
	\label{fig:otfsber}
\end{figure}

After presenting the important properties viz; OoB and PAPR, now we discuss BER performance  in EVA channel. Doppler is generated following Jake's spectrum, $\nu_p=\nu_{max} cos(\theta_p)$, where $\theta_p$ is uniformly distributed over $[-\pi ~ \pi]$ . The CP is chosen long enough to accommodate the wireless channel delay spread. We plot BER for different waveforms in Fig.~\ref{fig:otfsber}. It can be seen that the BER performance of CDPS-
OTFS is similar to that of RPS-OTFS. This can be ascribed to the unitary property of CDPS-OTFS. It can also be observed that both RPS-OTFS and CDPS-OTFS  have diversity gain of more than 10 dB over OFDM  at the BER of $10^{-3}$, which is aligned with results in \cite{raviteja_interference_2018,hadani_orthogonal_2017,surabhi_diversity_2019}. 

\section{Conclusion}
In this work, we proposed a circular Dirichlet pulse shaped OTFS (CDPS-OTFS), which can reduce OoB radiation significantly over rectangular pulse shaped OTFS (RPS-OTFS) system. CDPS-OTFS is also shown to have reduced PAPR over RPS-OTFS system. We have also developed a low complexity transmitter to implement CDPS-OTFS, which has log-linear complexity. This is achieved without any loss in BER performance and receiver complexity.  Due to these desirable properties, our proposed CDPS-OTFS can be attractive for practical high-speed vehicular communication systems.

\bibliographystyle{plain} 
\bibliography{GFDM_new,OTFS,5GScenariosRequirements,HighDopplerCommunication,manual,REF,precoded_GFDM,ComparisonofWaveforms,OFDM}

\end{document}